\documentclass[transaction]{IEEEtran}
\usepackage{graphicx}
\usepackage{graphicx}
\usepackage{epstopdf}
\usepackage[latin1]{inputenc}
\usepackage{amsmath,amssymb,amscd,latexsym,dsfont}
\usepackage[english]{babel}
\usepackage{tabularx,cite}
\usepackage{graphicx}
\usepackage{algpseudocode}
\usepackage{algorithm}
\usepackage{algorithmicx}
\usepackage{float}
\usepackage{multicol}
\usepackage{mathtools}
\usepackage{psfrag}
\usepackage{comment,psfrag,subfigure,enumerate}
\usepackage{color}
\usepackage{amsthm}
\ifCLASSINFOpdf
\else
\fi
\begin{document}
\title{Coordinated Hybrid Automatic Repeat Request; Extended Version}\vspace{-0mm}
\author{\IEEEauthorblockN{Behrooz Makki, Tommy Svensson, Thomas Eriksson and Mohamed-Slim Alouini, \emph{Fellow, IEEE}}\\\vspace{-0mm}
%\IEEEauthorblockA{Department of Signals and Systems,
%Chalmers University of Technology, Gothenburg, Sweden\\
%Email: \{behrooz.makki, alexandre.graell, thomase\}@chalmers.se}
%%\and
%%\IEEEauthorblockN{Thomas Eriksson}
%%\IEEEauthorblockA{Department of signals and systems\\
%%Chalmers University of Technology\\
%%Gothenburg, Sweden\\
%%Email: thomase@chalmers.se}
%\thanks{The authors are with the Department of ...}
\thanks{Behrooz Makki, Tommy Svensson and Thomas Eriksson are with Chalmers University of Technology, Email: \{behrooz.makki, tommy.svensson, thomase\}@chalmers.se. Mohamed-Slim Alouini is with the King Abdullah University of Science and Technology (KAUST), Email: slim.alouini@kaust.edu.sa}
}

%Telephone: (800) 555--1212, Fax: (888) 555--1212}
\maketitle
%\onecolumn
\vspace{-19mm}
\begin{abstract}
We develop a coordinated hybrid automatic repeat request (HARQ) approach. With the proposed scheme, if a user message is correctly decoded in the first HARQ rounds, its spectrum is allocated to other users, to improve the network outage probability and the users' fairness. The results, which are obtained for single- and multiple-antenna setups, demonstrate the efficiency of the proposed approach in different conditions. For instance, with a maximum of $M$ retransmissions and single transmit/receive antennas, the diversity gain of a user increases from $M$ to $(J+1)(M-1)+1$ where $J$ is the number of users helping that user.
%Considerable performance improvement is achieved even with a single retransmission round.
\end{abstract}
% \fi
\IEEEpeerreviewmaketitle
%
%\section{Introduction}
%
%
%
%The rest of the paper is organized as follows. System model and problem statement are illustrated in section II. Then, the theoretical concepts are presented in section III. Section IV presents some simulation results and, finally, the last section concludes the paper.
\vspace{-0mm}
%
%
%\linespread{1.65}
\section{Introduction}
Hybrid automatic repeat request (HARQ) is a well-established approach for reliable wireless communication \cite{throughputdef}.
%Utilizing both forward error correction and error detection, HARQ techniques reduce the data outage probability by retransmitting the data which has experienced \emph{bad} channel conditions.
There are many works improving the performance of HARQ protocols via optimal rate/power allocation, e.g., \cite{throughputdef,6006606,1661837,6235077}. On the other hand, the long-term evolution (LTE) standards provide the capability for dynamic resource allocation in the frequency domain \cite[Subsection 16.5.8]{LTEbook}. Thus, it is interesting to analyze the performance of HARQ protocols using dynamic frequency allocation.

This letter introduces a \emph{coordinated} HARQ approach. Here, the frequency resources are dynamically allocated among the users based on the HARQ feedback signals. The results are obtained for the repetition time diversity (RTD) and the incremental redundancy (INR) HARQ protocols utilizing single or multiple transmit/receive antennas. As demonstrated in the paper, the advantages of the proposed scheme are: 1) all users benefit from a substantial outage probability improvement and 2) the users' fairness is improved considerably. This is of interest because the fairness has been investigated only in a few HARQ-based systems, e.g., \cite{6595163,5371591}. Moreover, 3) the proposed coordinated approach is useful for buffer-limited transmitters. In harmony with all fairness-based schemes, the coordination may reduce the throughput of the users with the best average channel characteristics slightly. However, the throughput degradation is very limited, as seen in the sequel. Finally, the coordination scales up the diversity gain of the users substantially.

The problem setup of the paper is different from \cite{throughputdef,6006606,1661837} (resp. \cite{6235077}) that optimize the performance of single-user (resp. cognitive radio) systems via rate/power adaptation in power-limited (resp. interference-limited) conditions. Also, we investigate a different problem from \cite{6595163} (resp. \cite{5371591}) which analyzes the fairness-adaptive throughput optimization in HARQ-based systems using adaptive modulation (resp. the fairness in relay-HARQ systems). Finally, our discussions on the users' message decoding probabilities, the diversity gain and the fairness have not been presented before\footnote{The uploaded file is an extended version of \cite{coordinatedH}.}.
%The analytical and numerical results show that the proposed approach leads to considerable outage probability and users' fairness improvement, with limited throughput degradation. Also, the coordination scales up the diversity gain of the users substantially.

%ARQ is a technique in the \emph{data link layer} already provided in many wireless protocols, e.g., IEEE 802.11n \cite{ieeestandard1} and IEEE 802.16e \cite{ieeestandard2}. Hence, it needs no additional design which introduces it as a cost- and complexity-efficient approach.
\vspace{-2mm}

%In the theoretical analysis, the ARQ protocols are normally designed to operate in fixed frequency resources. However, the practical standards have provided capability to improve the performance of ARQ protocols by utilize different frequency bands in the retransmissions.
\section{System model}
Consider an HARQ protocol with a maximum of $M$ retransmissions.
%, i.e., the data is (re)transmitted a maximum of $M$ times.
Also, define a packet as the transmission of a codeword along with all its possible retransmissions and let $P$ be the transmission power for each frequency band. We study block-fading conditions where the channel coefficients remain constant in each retransmission and then change to other values based on their probability density functions (pdf:s). At time slot $t$, the channel coefficient associated with the $i$th frequency band is represented by $\prescript{}{i}{h}(t)$ and we define $\prescript{}{i}g(t)\doteq|\prescript{}{i}h(t)|^2$ which is referred to as the channel gain. For Rayleigh-fading channels, on which we focus, the channel gains follow the pdfs
%$h_{\text{A}}\sim\mathcal{CN}(0,\frac{1}{\lambda_{\text{A}}}),\, h_{\text{B}}\sim\mathcal{CN}(0,\frac{1}{\lambda_{\text{B}}})$. Thus, the gains pdf are given by
${f_{{\prescript{}{i}g}}}(x) = {\prescript{}{i}\lambda}{e^{ - {\prescript{}{i}\lambda}x}},x \ge 0,$ where $\prescript{}{i}\lambda$ is the fading parameter of the $i$th channel. In each link, the channel coefficient is assumed to be known by the receiver.
%, which is an acceptable assumption in block-fading channels \cite{throughputdef,6006606,tuninetti2011,1661837,6235077}.
However, there is no instantaneous channel state information available at the transmitters.
%\footnote{The transmitter can be considered as an orthogonal frequency-division multiplexing (OFDM)-based base station (BS) or the BSs in a coordinated multi-point (CoMP) network with perfect backhaul links.} except the HARQ feedback bits.
%The results are obtained for two different HARQ protocols:
%\begin{itemize}
%\item \textbf{RTD.} Here, the same data is repeated in the (re)transmission rounds and, in each round, the receiver performs maximum ratio combining (MRC) of all received signals.
%\item \textbf{INR.} With this scheme, a codeword is sent with an aggressive rate in the first round. Then, if the receiver cannot decode the codeword, further parity bits are sent in the next retransmission rounds and in each round the receiver decodes the data based on all received signals.
%\end{itemize}

\emph{Coordination Model:} The transmitter has access to $K$ frequency bands each having normalized bandwidth $W=1$ (it is straightforward to extend the results to the cases with different bandwidths).
%\footnote{Extension of results to the case with different bandwidths is straightforward.}
%For simplicity, we start the discussions by considering the special case of $K=2$ and $M=2,$ which illustrates the main aspects of the proposed scheme. Extension of the results to the cases with $K\ge 2$ and $M\ge 2$ is investigated in Subsections III.C-E.
The data transmission protocol is designed as follows. With $K$ users, separate frequency bands are first allocated to the users. If none of the users can correctly decode their corresponding codewords (resp. all users correctly decode their corresponding messages), there is no coordination between the frequency resources, and each user receives its corresponding retransmission (resp. a new message transmission) in the next time slot. The frequency coordination occurs if some of the users successfully decode their corresponding codewords, while the other users cannot. In this case, all frequency resources of the next slot are allocated to the users with unsuccessful message decoding, for which the messages are retransmitted. Denoting the complement of the event $s$ by $\bar s$ and $A_nB_m$ as the event that users A and B correctly decode their corresponding messages in rounds $n$ and $m$, respectively, the following example demonstrates the data transmission protocol for the simplest case with $K=2$ and $M=2$ (Also, an illustrative example of the cooperation approach is given in Fig.3 at the end of the paper).

\textbf{Example:} Start the data transmission by sending separate messages to users A and B. The following cases may occur in the next time slot:
\begin{itemize}
  \item If both users correctly decode their corresponding messages, represented by the event $A_1B_1,$ a new packet transmission starts for each user, within its associated frequency band.
  \item If none of the users decode its corresponding message, shown by the event $\bar A_1\bar B_1,$ the data is retransmitted for each user, within its associated frequency band.
  \item If user A correctly decodes its message while user B cannot, represented by $A_1\bar B_1,$ both frequency bands of the next slot are allocated to user B. That is, in the next time slot the codeword of user B is retransmitted in two frequency bands, which is the same as two \emph{simultaneous} retransmissions. Finally, the same procedure is considered if user B successfully decodes the message in round 1, while user A cannot.
\end{itemize}
\vspace{-3mm}
\section{Analysis}
In this section, we analyze the users' outage probability and the system throughput.
%for the considered data transmission scheme.
For simplicity, we first concentrate on the special case of $M=K=2$ with single-antenna transmission. Later, the results are extended to the cases with $M\ge 2$, $K\ge 2$ and multiple-input-multiple-output (MIMO) transmission. We study the system performance for the RTD and the INR HARQ protocols as two efficient schemes leading to high throughput and low outage probability \cite{throughputdef,6006606,1661837,6235077}. Straightforward modifications can be applied for the cases with basic HARQ.

%Let $P$ be the transmission power for each frequency band. Also, at time slot $t$, the channel coefficient associated with the $i$th frequency band is represented by $\prescript{}{i}{h}(t)$ and we define $\prescript{}{i}g(t)\doteq|\prescript{}{i}h(t)|^2$ which is referred to as the channel gain.
%$Q_\text{A}$ and $Q_\text{B}$ information nats are considered for each packet transmission of users A and B, respectively.
%, where a packet is defined as the transmission of a codeword along with all its possible retransmission rounds.
%The system performance in the presence of the RTD and the INR HARQ protocols is studied as follows.
\vspace{-3mm}
\subsection{RTD Protocol}
Using RTD with codewords of length $L$, $Q_\text{A}$ and $Q_\text{B}$ information nats are encoded in each codeword of users A and B, respectively. Thus, the initial transmission rates are $R_\text{A}\doteq\frac{Q_\text{A}}{L},\,R_\text{B}\doteq\frac{Q_\text{B}}{L}$. For each user, the same codeword is retransmitted in the successive retransmission rounds and the receiver performs maximum ratio combining (MRC) of all received signals \cite{throughputdef}. Hence, the equivalent transmission rates after $m$ retransmissions are $R_{(m),\text{A}}=\frac{Q_\text{A}}{mL}=\frac{R_\text{A}}{m}$ and $R_{(m),\text{B}}=\frac{Q_\text{B}}{mL}=\frac{R_\text{B}}{m}$. Utilizing the first frequency band, the data transmission of user A stops at the end of the first round if
$\log(1+\prescript{}{1}g(t)P)>R_\text{A},$ otherwise the codeword is retransmitted. Thus, with $M=2,$ different events may occur in each time slot, whose probabilities are given by
%the probabilities of being in states $\{\text{A}_1\text{B}_1\},\, \{\text{A}_2\text{B}_2\},\, \{\text{A}_2 \text{A}_2\}$ and $\{\text{B}_2\text{B}_2\}$ are found by
\vspace{-1mm}
\begin{align}\label{eq:equationasli1}
&\Pr(A_2B_2)+\Pr(\bar A_2B_2)+\Pr(A_2\bar B_2)+\Pr(\bar A_2\bar B_2)=\alpha\beta\gamma,\nonumber
\\
&\Pr(A_2B_1)+\Pr(\bar A_2B_1)=\Pr(\bar A_1B_1)=\alpha(1-\beta)\gamma,\nonumber
\\
&\Pr(A_1B_2)+\Pr(A_1\bar B_2)=\Pr(A_1\bar B_1)=(1-\alpha)\beta\gamma,\nonumber
\\
&\Pr(A_1 B_1)=(1-\alpha)(1-\beta)\gamma,\nonumber
\\
&\gamma\doteq\Pr(A_1B_1)+\Pr(\bar A_1B_1)+\Pr(A_1\bar B_1)+\Pr(\bar A_1\bar B_1),\nonumber
\\
&\alpha\doteq\Pr(\log(1+\prescript{}{1}g(t)P)<R_\text{A}),\nonumber
\\
&\beta\doteq\Pr(\log(1+\prescript{}{2}g(t)P)<R_\text{B}).
\end{align}
%Here, e.g., $\Pr(B_2B_2)$ is the probability that user B correctly decodes its massage in round 2, while user A (resp. user B) could (resp. could not) decode its message in round 1.
Setting the sum of all possible probabilities equal to 1, the sum probability of all possible events in the first slot of the new packet transmissions, i.e., $\gamma$ in (\ref{eq:equationasli1}), is found as
\vspace{-2mm}
\begin{align}\label{eq:PAB11}
&\gamma=\frac{1}{1+\alpha+\beta-\alpha\beta},\vspace{-2mm}
\end{align}
from which the probabilities $\Pr(A_1B_1),$ $\Pr(A_1\bar B_1),$ $\Pr(\bar A_1\bar B_1)$ and $\Pr(\bar A_1B_1)$ are obtained (see (\ref{eq:equationasli1})).

Given that user A successfully decodes its corresponding message at the end of round 1 while user B cannot decode its associated codeword, two copies of the user B's codeword are retransmitted in the two frequency bands of the next slot. The receiver of user B performs MRC of the three received signals (1 transmission plus 2 retransmissions). Hence,
%the received signal-to-noise ratio (SNR) of user B at the end of the second round increases to $\text{SNR}_{B}^{(2)}=P(g_\text{B}(t)+g_\text{B}(t+1)+g_\text{A}(t+1))$ and
we have
\vspace{-2mm}
\begin{align}\label{eq:probBBbars2}
&\Pr(A_1\bar B_2)=\gamma\Pr\Big(\log(1+\prescript{}{1}g(t)P)\ge R_\text{A} \,\,\cap  \nonumber\\& \log\big(1+\big(\prescript{}{2}g(t)+\prescript{}{2}g(t+1)+\prescript{}{1}g(t+1)\big)P\big)<R_\text{B} \Big),
\end{align}
%\vspace{-4mm}
%\begin{align}\label{eq:probBBs2}
%&\Pr(B_2B_2)=(1-\alpha)\beta\gamma-\Pr(\bar B_2\bar B_2),
%\end{align}
%Here, $\Pr(\{{\text{B}_2\text{B}_2}\}^{\{s\}})$ denotes the probability of being in state $\{\text{B}_2\text{B}_2\}$ and successfully decoding the massage of user B (while it was not decoded before). Note that in (\ref{eq:probBBs2})-(\ref{eq:probBBbars2})
and $\Pr(A_1B_2)=(1-\alpha)\beta\gamma-\Pr(A_1\bar B_2).$ Here, we have used the fact that with the signal-to-noise ratio (SNR) of $\text{SNR}_i$ for the $i$th received signal the maximum decodable rate is
\vspace{-2mm}
\begin{align}\label{eq:achievablerateRTD}
U_{(m)}^{\text{RTD}}=\frac{1}{m}\log\left(1+\sum_{i=1}^{m}\text{SNR}_i\right),
\end{align}
if the same codeword is retransmitted $m$ times \cite[Section III]{throughputdef}.

For Rayleigh-fading channels,
%on which we focus, the channel gains follow the pdf:s
%$h_{\text{A}}\sim\mathcal{CN}(0,\frac{1}{\lambda_{\text{A}}}),\, h_{\text{B}}\sim\mathcal{CN}(0,\frac{1}{\lambda_{\text{B}}})$. Thus, the gains pdf are given by
%${f_{{\prescript{}{i}g}}}(x) = {\prescript{}{i}\lambda}{e^{ - {\prescript{}{i}\lambda}x}},x \ge 0,$ where $\prescript{}{i}\lambda$ is the fading parameter of the $i$th channel.
% determined based on the path loss and shadowing between the terminals.
%Thus,
(\ref{eq:probBBbars2}) is found as\footnote{The analytical results are given for $\prescript{}{i}\lambda\ne\prescript{}{j}\lambda, i\ne j.$ Straightforward modifications should be applied for the cases with $\prescript{}{i}\lambda=\prescript{}{j}\lambda, i\ne j$.}
\vspace{-1mm}
\begin{align}\label{eq:pdfprobBBbars2}
&\Pr(A_1\bar B_2)=\gamma(1-\alpha)\Phi,\,\,\alpha=1-e^{-\prescript{}{1}\lambda C_\text{A}},\nonumber\\&\Phi=\Pr(\prescript{}{2}g(t)+\prescript{}{2}g(t+1)+\prescript{}{1}g(t+1)<C_\text{B})\nonumber\\&=\int_0^{C_\text{B}}{\int_0^{C_\text{B}-x}f_{\prescript{}{1}g}(x)f_{\prescript{}{2}g}(y)}\Pr(\prescript{}{2}g(t+1)<C_\text{B}-x-y)\text{d}x\text{d}y
\nonumber\\&=\int_0^{C_\text{B}}\int_0^{C_\text{B}-x}\prescript{}{1}\lambda e^{-\prescript{}{1}\lambda x}\prescript{}{2}\lambda e^{-\prescript{}{2}\lambda y}(1-e^{-\prescript{}{2}\lambda (C_\text{B}-x-y)})\text{d}x\text{d}y \nonumber\\&=1-e^{-\prescript{}{1}\lambda C_\text{B}}+\frac{e^{-\prescript{}{2}\lambda C_\text{B}}-e^{-\prescript{}{1}\lambda C_\text{B}}}{\frac{\prescript{}{2}\lambda}{\prescript{}{1}\lambda}-1}+\frac{C_\text{B}e^{-\prescript{}{2}\lambda C_\text{B}}}{\frac{1}{\prescript{}{1}\lambda}-\frac{1}{\prescript{}{2}\lambda}}\nonumber\\&+\frac{\prescript{}{1}\lambda \prescript{}{2}\lambda}{(\prescript{}{1}\lambda-\prescript{}{2}\lambda)^2}(e^{-\prescript{}{2}\lambda C_\text{B}}-e^{-\prescript{}{1}\lambda C_\text{B}}),
\end{align}
where $C_\text{A}\doteq \frac{e^{R_\text{A}}-1}{P},$ $C_\text{B}\doteq \frac{e^{R_\text{B}}-1}{P}.$ The other probabilities, e.g., $\Pr(\bar A_2B_1), \Pr(A_2B_1),$
%\vspace{-0mm}
%\begin{align}
%&\Pr(\bar A_2B_1)=\gamma\Pr\Big(\log(1+\prescript{}{2}g(t)P)\ge R_\text{B} \,\cap  \nonumber\\& \log\big(1+\big(\prescript{}{1}g(t)+\prescript{}{1}g(t+1)+\prescript{}{2}g(t+1)\big)P\big)<R_\text{A} \Big),
%\end{align}
%\vspace{-6mm}
%\begin{align}\label{eq:probAAs22}
%&\Pr(A_2B_1)=\alpha(1-\beta)\gamma- \Pr(\bar A_2B_1),
%\end{align}
$\Pr(A_1B_2)$ and the outage probabilities, e.g., $\Pr(\text{Outage}_\text{B})=\Pr(A_2\bar B_2)+\Pr(\bar A_2\bar B_2)+\Pr(A_1\bar B_2)$ are found with the same procedure as in (\ref{eq:pdfprobBBbars2}) leading to
%. In this way, following the same arguments as in (\ref{eq:pdfprobBBbars2}), the outage probability of, e.g., user B is obtained as
\vspace{-1mm}
\begin{align}
&\Pr(\text{Outage}_\text{B})=\gamma\alpha(1-e^{-\prescript{}{2}\lambda C_\text{B}}-\prescript{}{2}\lambda C_\text{B}e^{-\prescript{}{2}\lambda C_\text{B}})\nonumber\\&+\gamma(1-\alpha)\Big(1-e^{-\prescript{}{1}\lambda C_\text{B}}+\frac{e^{-\prescript{}{2}\lambda C_\text{B}}-e^{-\prescript{}{1}\lambda C_\text{B}}}{\frac{\prescript{}{2}\lambda }{\prescript{}{1}\lambda}-1}+\frac{C_\text{B}e^{-\prescript{}{2}\lambda C_\text{B}}}{\frac{1}{\prescript{}{1}\lambda }-\frac{1}{\prescript{}{2}\lambda}}\nonumber\\&+\frac{\prescript{}{1}\lambda \prescript{}{2}\lambda}{(\prescript{}{1}\lambda-\prescript{}{2}\lambda)^2}(e^{-\prescript{}{2}\lambda C_\text{B}}-e^{-\prescript{}{1}\lambda C_\text{B}})\Big).
\end{align}

The throughput (in nats-per-channel-use (npcu)) is defined as
\vspace{-1mm}
\begin{align}
&\eta=\lim_{N\to \infty}\frac{\sum_{t=1}^N{\tilde Q_\text{A}(t)}+\sum_{t=1}^N{\tilde Q_\text{B}(t)}}{NL},
\end{align}
where $\sum_{t=1}^N{\tilde Q_\text{A}(t)}$ and $\sum_{t=1}^N{\tilde Q_\text{B}(t)}$ denote the total number of information nats that are successfully decoded by users A and B, respectively, in $N$ time slots \cite{throughputdef}. Using (\ref{eq:equationasli1})-(\ref{eq:pdfprobBBbars2}), the law of large numbers and $N\to\infty$ time slots, the total number of information nats successfully decoded by user A is found as
\vspace{-0mm}
\begin{align}
\sum_{t=1}^N{\tilde Q_\text{A}(t)}&=Q_\text{A}N\Big(\Pr(A_1\bar B_1)+\Pr(A_1B_1)\nonumber\\&+\Pr(A_2B_1)+\Pr(A_2B_2)+\Pr(A_2\bar B_2)\Big).\nonumber
\end{align}
Thus, from $R_\text{A}=\frac{Q_\text{A}}{L},\,R_\text{B}=\frac{Q_\text{B}}{L},$ the throughput is obtained by
\begin{align}\label{eq:etaakhar}
&\eta=R_\text{A}\Big(\Pr(A_1\bar B_1)+\Pr(A_1B_1)+\Pr(A_2B_1)+\Pr(A_2B_2)\nonumber\\&+\Pr(A_2\bar B_2)\Big)+R_\text{B}\Big(\Pr(\bar A_1B_1)+\Pr(A_1B_1)+\Pr(A_1B_2)\nonumber\\&+\Pr(A_2B_2)+\Pr(\bar A_2B_2)\Big).
\end{align}
\vspace{-9mm}
\subsection{INR Protocol}
Using INR, new codewords are sent in the successive retransmission rounds
% of a packet. Also, in each (re)transmission round
and the message is decoded by the receivers using all previously received signals of the packet. In this case, the results of \cite{throughputdef,6006606,1661837,6235077} can be used to rephrase the INR-based probability terms as, e.g.,
\vspace{-2mm}
\begin{align}\label{eq:inrprobBBs2}
&\Pr(A_1B_2)=\gamma\Pr\Big(\log(1+\prescript{}{1}g(t)P)\ge R_\text{A} \,\cap\nonumber\\& \log(1+\prescript{}{2}g(t)P)<R_\text{B}\le  \log(1+\prescript{}{2}g(t)P)\nonumber\\&+\log(1+\prescript{}{2}g(t+1)P)+\log(1+\prescript{}{1}g(t+1)P) \Big).
\end{align}
That is, the achievable rate terms $U_{(m)}^{\text{RTD}}=\frac{1}{m}\log(1+\sum_{i=1}^{m}{\text{SNR}_i})$ of the RTD, i.e., (\ref{eq:achievablerateRTD}), are replaced by the terms
\vspace{-2mm}
\begin{align}\label{eq:rateinr}
U_{(m)}^{\text{INR}}=\frac{1}{m}\sum_{i=1}^{m}\log(1+{\text{SNR}_i})
\end{align}
in the INR, and the probabilities are recalculated. This is the only modification required for the INR, compared to the RTD, and the rest of the discussions remain the same as before.
\vspace{-3mm}
\subsection{Extension of Results to Arbitrary Number of Retransmissions}
The results can be extended to the case with a maximum of $M\ge 2$ retransmissions. Here, the probability that, for instance, users A and B successfully decode their messages at the $n$th and $m$th, $n\le m,$ rounds of the RTD, respectively, is obtained by
\vspace{-1mm}
\begin{align}\label{eq:probABnm}
\begin{array}{l}
\Pr (A_nB_m) = \gamma \Pr \Big(\log (1 + P\sum_{i = 0}^{n - 2} {{\prescript{}{1}g}(t + i)} ) <\\ R_\text{A} \le
\log \big(1 + P\sum\limits_{i = 0}^{n - 1} {{\prescript{}{1}g}(t + i)} \big) \,\cap \\
\log (1 + P\sum\limits_{i = 0}^{m - 2} {{\prescript{}{2}g}} (t + i) + P\sum\limits_{i = n}^{m - 2} {{\prescript{}{1}g}} (t + i)) < R_\text{B} \le \\
\log (1 + P\sum\limits_{i = 0}^{m - 1} {{\prescript{}{2}g}} (t + i) + P\sum\limits_{i = n}^{m - 1} {{\prescript{}{1}g}} (t + i))\Big), n \le m,
\end{array}
\end{align}
and the other terms, e.g., $\eta,$ $\Pr(\text{Outage}_\text{A})$ and $\Pr(\text{Outage}_\text{B})$ are rephrased correspondingly.

In (\ref{eq:pdfprobBBbars2}), we presented a closed-form expression for the probabilities, e.g., $\Pr(A_1\bar B_2),$ with $M=2$. Theorem 1 extends the results to the cases with arbitrary number of retransmissions.

\textbf{Theorem 1}: For Rayleigh fading channels, the throughput and the outage probability of the proposed RTD- and INR-based schemes are obtained via the following equalities, respectively
\vspace{-4mm}
\begin{align}\label{eq:theoremeq11}
\begin{array}{l}
\Pr(\log(1+P(\sum_{k=0}^{n-1}{{\prescript{}{1}g(t+k)}}+\sum_{l=t'}^{t'+m-1}{{\prescript{}{2}g(t+l)}}))<x)\\=\mathcal{W}(e^x-1)-\mathcal{W}(0),\\
\mathcal{W}(x)\doteq-\sum_{k=1}^{n}{\frac{a_k\Gamma(k,\prescript{}{1}\lambda x)}{(k-1)!}}-\sum_{k=1}^{m}{\frac{b_k\Gamma(k,\prescript{}{2}\lambda x)}{(k-1)!}},\\a_k\doteq(-\frac{\prescript{}{1}\lambda}{\prescript{}{2}\lambda})^{n-k}{{n+m-k-1}\choose {n-k}}(1-\frac{\prescript{}{1}\lambda}{\prescript{}{2}\lambda})^{-(n+m-k)},\\
b_k\doteq(-\frac{\prescript{}{2}\lambda}{\prescript{}{1}\lambda})^{m-k}{{n+m-k-1}\choose {m-k}}(1-\frac{\prescript{}{2}\lambda}{\prescript{}{1}\lambda})^{-(m+n-k)},
\end{array}
\end{align}
\vspace{-6mm}
\begin{align}\label{eq:theorem12}
\begin{array}{l}
\Pr(\sum\limits_{k=0}^{n-1}{\log(1+{\prescript{}{1}g(t+k)}P})+ \sum\limits_{l=t'}^{t'+m-1}{\log(1+{\prescript{}{2}g(t+l)}P}) <x)\\= 1-{e^{\frac{ \prescript{}{1}\lambda n+\prescript{}{2}\lambda m}{P}}}\times\\ \mathcal{Y}_{n+m+1,1}^{n+m+1,0}\bigg[\frac{\prescript{}{1}\lambda^n \prescript{}{2}\lambda^m }{P^{n+m}}e^{x}\bigg |_{(0,1,0),\underbrace{(1,1,\frac{\prescript{}{1}\lambda}{P}),\ldots,(1,1,\frac{\prescript{}{1}\lambda}{P})}_{n \text{ times}},\dots}^{\,\,\,\,\,\,\,\,\,\,\,\,\,\,\,\,\,\,\,\,\,\,\,\,\,\,\,\,\,\,\,\,\,\,\,\,\,\,(1,1,0)} \\\,\,\,\,\,\,\,\,\,\,\,\,\,\,\,\,\,\,\,\,\,\,\,\,\,\,\,\,\,\,\,\,\,\,\,\,\,\,\,\,\,\,\,\,\,\,\,\,\,\,\,\,\,\,\,\,\,\,\,\,\,\,_{\ldots,\underbrace{(1,1,\frac{\prescript{}{2}\lambda}{P}),\ldots,(1,1,\frac{\prescript{}{2}\lambda}{P})}_{m \text{ times}}}^{}\bigg] \forall t',n,m.
\end{array}
\end{align}
Here, $\Gamma(.,.)$ and $\mathcal{Y}_{s_1,s_2}^{s_3,s_4}[.|_.^.]$ are the incomplete Gamma function and the generalized upper incomplete Fox'H function \cite{5357980}, respectively. Also, $n\choose k$ denotes the ``\emph{n choose k}'' operator.
\vspace{-2mm}
\begin{proof}
%See \cite[Theorem 1]{extendedcoord}.
Following (\ref{eq:probBBbars2})-(\ref{eq:probABnm}) and the same discussions as in \cite[Section IV]{throughputdef}, it is straightforward to show that the throughput and the outage probability of the RTD- and INR-based schemes can be represented as monotonic functions of the probabilities
\vspace{-1mm}
\begin{align}
\left\{ \begin{array}{l}
\pi_\text{RTD}=\Pr \big(\log \big(1 + P(\sum_{k = 0}^{n - 1} {{\prescript{}{1}g}(t + k)} \\\,\,\,\,\,\,\,\,\,\,\,\,\,\,\,\,\,\,
 + \sum_{l = t'}^{t' + m - 1} {{\prescript{}{2}g}(t + l)} )\big) < x\big),\,\,\,\,\,\,\,\,\,\,\,\,\,\,\,\,\,\,\,\,\,\,\,\,\,\text{           For RTD}\\
\pi_\text{INR}=\Pr \big(\sum_{k = 0}^{n - 1} {\log (1 + {\prescript{}{1}g}(t + k)P} )  \\\,\,\,\,\,\,\,\,\,\,\,\,\,\,\,\,\,\,
+\sum_{l = t'}^{t' + m - 1} {\log (1 + {\prescript{}{2}g}(t + l)P} ) < x\big),\,\,\,\,\text{    For INR.}
\end{array} \right.\nonumber
\end{align}
To find $\pi_\text{RTD}$ (and then the throughput and outage probability), we use Laplace transform $\mathcal{L}\{.\}$ and its inverse $\mathcal{L}^{-1}\{.\}$ to write
\vspace{-2mm}
\begin{align}
\begin{array}{l}
\pi_\text{RTD}\mathop  = \limits^{(a)}\int_0^{e^x-1}{   \mathcal{L}^{-1}\{\frac{1}{(1+\frac{Ps}{\prescript{}{1}\lambda})^{n}{(1+\frac{Ps}{\prescript{}{2}\lambda})^{m}}}\}\text{d}z}\\\mathop  = \limits^{(b)}{\int_0^{e^x-1}{\mathcal{L}^{-1}\{\sum_{k=1}^{n}{\frac{a_k}{(1+\frac{Ps}{\prescript{}{1}\lambda})^{k}}}+\sum_{k=1}^{m}{\frac{b_k}{(1+\frac{Ps}{\prescript{}{2}\lambda})^{k}}}\}\text{d}z}   }\\\mathop  = \limits^{(c)}\int_{0}^{e^x-1}{(\sum_{k=1}^{n}{\frac{a_k \prescript{}{1}\lambda^kz^{k-1}e^{-\frac{\prescript{}{1}\lambda z}{P}}}{(k-1)!}}+\sum_{k=1}^{m}{\frac{b_k \prescript{}{2}\lambda^kz^{k-1}e^{-\frac{\prescript{}{1}\lambda z}{P}}}{(k-1)!}})\text{d}z}\\=\mathcal{W}(e^x-1)-\mathcal{W}(0).\nonumber
\end{array}
\end{align}
Here, $(a)$ follows from the fact that the pdf of the sum of independent random variables is obtained by the convolution of their pdfs and $\mathcal{L}\{f_{\prescript{}{i}\theta}\}=(1+\frac{Ps}{\prescript{}{i}\lambda})^{-1},\prescript{}{i}\theta\doteq \prescript{}{i}gP.$ Then, $(b)$ is obtained by partial fraction of ${(1+\frac{Ps}{\prescript{}{1}\lambda})^{-n}{(1+\frac{Ps}{\prescript{}{2}\lambda})^{-m}}}$, with fraction coefficients $a_k,b_k$ given in (\ref{eq:theoremeq11}). Also, $(c)$ is derived by inverse Laplace transform and some manipulations.

Finally, the probabilities $\pi_\text{INR}$ of the INR are obtained by appropriate parameter setting in \cite[eq. 18]{5357980} leading to (\ref{eq:theorem12}). Then, having $\pi_\text{RTD}$, $\pi_\text{INR},$ the system performance is analyzed with the same method as in (\ref{eq:equationasli1})-(\ref{eq:probABnm}).
%, as stated in the theorem.
\end{proof}
Note that the generalized upper incomplete Fox'H function has an efficient MATHEMATICA implementation \cite{5357980}.
\vspace{-3mm}
\subsection{Multiple-Antenna Scenario}
From another perspective, we can extend the results to the case with MIMO transmission; consider a setup with $u$ transmit antennas and $v$ receive antennas for each user. Let $\prescript{}{i}H(t)\in \mathcal{C}^{v\times u}$ denote the complex channel matrix associated with the $i$th frequency band at time slot $t$. Also, represent the $u\times u$ identity matrix by $I_{u}$. Using isotropic input distribution over all transmit antennas, the same procedure as in \cite[Section III.C]{6006606} can be used to rephrase
%
%\begin{align}
%\textbf{P}=\frac{P}{u}\textbf{I}_{u},
%\end{align}
the achievable rate term of the RTD, i.e., (\ref{eq:achievablerateRTD}), as
\vspace{-2mm}
\begin{align}
&U_{(n,m),\text{B}}^{\text{RTD}}=\frac{1}{2m-n}\log\det(I_{(2m-n)v}+\frac{P}{u}{H}_{(n,m),\text{B}}^{\text{RTD}}({H}_{(n,m),\text{B}}^{\text{RTD}})^*),\nonumber\\&{H}_{(n,m),\text{B}}^{\text{RTD}}=\nonumber\\&\,\,{[ {{\prescript{}{2}H^\text{T}(t)}\ldots{\prescript{}{2}H^\text{T}(t+m-1)}}  \, {{\prescript{}{1}H^\text{T}(t+n)}\ldots{\prescript{}{1}H^\text{T}(t+m-1)}}]^\text{T}},\nonumber
\end{align}
if the data retransmissions of users A and B continue up to the end of rounds $n$ and $m$, $n\le m$, respectively. Here, $\det({X}),$ ${X}^\text{T}$ and ${X}^*$ represent the determinant, the transpose and the Hermitian of the matrix ${X}$, respectively. For INR, we have
\vspace{-2mm}
\begin{align}\label{eq:INRMIMO}
&U_{(n,m),\text{B}}^{\text{INR}}=\frac{1}{2m-n}\Big(\sum_{i=0}^{m-1}\log\det\big({I}_{v}+\frac{P}{u}\prescript{}{2}H(t+i)\prescript{}{2}H(t+i)^*\big)\nonumber\\&+\sum_{i=n}^{m-1}\log\det\big({I}_{v}+\frac{P}{u}\prescript{}{1}H(t+i)\prescript{}{1}H(t+i)^*\big)\Big).
\end{align}
In this way, using
\begin{align}\label{eq:rtdMIMO2}
&U_{(n,0),\text{A}}^{\text{RTD}}\doteq \frac{1}{n}\log\det\bigg({I}_{nv}+\frac{P}{u}H_{(n,0),\text{A}}^{\text{RTD}}(H_{(n,0),\text{A}}^{\text{RTD}})^*\bigg),\nonumber\\& H_{(n,0),\text{A}}^{\text{RTD}}=[{{\prescript{}{1}H^\text{T}(t)}\ldots{\prescript{}{1}H^\text{T}(t+n-1)}}]^\text{T},
\end{align}
the probabilities, e.g., (\ref{eq:probABnm}), are obtained by
\begin{align}\label{eq:rtdMIMO3}
&\Pr(A_nB_m)=\Pr\Big(U_{(n-1,0),\text{A}}^{\text{RTD}}<\frac{R_\text{A}}{n-1}\,\cap\nonumber\\& U_{(n,0),\text{A}}^{\text{RTD}}\ge\frac{R_\text{A}}{n}\,\cap U_{(n,m-1),\text{B}}^{\text{RTD}}<\frac{R_\text{B}}{2(m-1)-n}\,\cap \nonumber\\&U_{(n,m),\text{B}}^{\text{RTD}}\ge\frac{R_\text{B}}{2m-n}\Big),
\end{align}
for RTD, while the rest of the arguments remain the same as in Subsection III.A. Finally, we can use (\ref{eq:INRMIMO}) and the same procedure as in (\ref{eq:rtdMIMO2})-(\ref{eq:rtdMIMO3}) to derive the probabilities for the INR.
\vspace{-4mm}
\subsection{Coordination with $K> 2$ Users}
The system performance in the presence of $K>2$ users depends on the designed coordination rules. However, Theorem 2 shows that assigning the free frequency bands of $J$ users to a user scales up its diversity gain, i.e., the negative of the slope of its outage probability curve at high SNRs, to $d=(J+1)(M-1)+1$.
%considering arbitrary number of users, the following theorem demonstrates substantial diversity gain increment achieved by frequency coordination.

\textbf{Theorem 2:} Using INR, the diversity gain of a user is $d=(J+1)(M-1)+1,$ if the coordination rule can provide it with the free frequency bands of $J$ users.
\begin{proof}
%See \cite[Theorem 2]{extendedcoord}.
With no loss of generality, let us consider the $K$th (the last) user and assume that it can utilize the free frequency resources of the first $J$ users. The diversity gain $d_K=-\lim_{P\to \infty}{\frac{\log(\Pr(\text{Outage}_K))}{\log P}}$ \cite[eq. 14]{1661837} of user $K$ is found as
\vspace{-2mm}
\begin{align}
\begin{array}{l}
{d_K} =  - \mathop {\lim }\limits_{P \to \infty } \frac{{\log \big(\Pr \big(\mathop  \cup \limits_{\forall {n_j} = 1, \ldots ,M,j = 1,\ldots,J} \xi_K ({n_1},\ldots,{n_{J}})\big)\big)}}{{\log P}} \\ \mathop  = \limits^{(a)}  - \mathop {\lim }\limits_{P \to \infty } \frac{{\log \big(\Pr \big(\big(\mathop  \cap \limits_{j = 1,\ldots,J} \omega_j ({n_j})\big) \cap \phi_K ({n_1},\ldots,{n_{J}})\big)\big)}}{{\log P}}
 \\ \mathop  = \limits^{(b)}  - \mathop {\lim }\limits_{P \to \infty } \frac{{\log \big(\prod_{j = 1}^{J} {({P^{ - ({n_j} - 1)}} - {P^{ - {n_j}}})} {P^{ - \big(M + \sum_{j = 1}^{J} {\left(M - {n_j}\right)} \big)}}\big)}}{{\log P}} \\=  - \mathop {\lim }\limits_{P \to \infty } \frac{{\log ({P^{ - (\sum_{j = 1}^{J} {({n_j} - 1)}  + M + \sum_{j = 1}^{J} {(M - {n_j})} )}})}}{{\log P}} \\= (J + 1)(M - 1) + 1,\\\xi_K ({n_1},\ldots,{n_{J}}) \buildrel\textstyle.\over= \{ \text{Outage}_K\,\& \,{c_j} = {n_j}\}, \\
\omega_j ({n_j}) \buildrel\textstyle.\over= \{ \sum\limits_{t = 1}^{{n_j} - 1} {\log (1 + {\prescript{}{j}g}(t)P)}  < R \le \sum\limits_{t = 1}^{{n_j}} {\log (1 + {\prescript{}{j}g}(t)P)} \} \\
\phi_K ({n_1},\ldots,{n_{J}}) \buildrel\textstyle.\over= \\\{ \sum\limits_{t = 1}^M {\log (1 + {\prescript{}{K}g}(t)P)}  + \sum\limits_{j = 1}^{J} {\sum\limits_{t = {n_j} + 1}^M {\log (1 + {\prescript{}{j}g}(t)P)} }  < R\}.\nonumber
\end{array}
\end{align}
Here,
%$g_k(t)$ represents the channel gain of user $k$ at slot $t$. Also,
%$x \equiv y$ is defined as $\mathop {\lim }\limits_{P \to \infty } \frac{x}{y} = 1$ and
$c_j$ is the indicator of the slot number in which the $j$th user message is decoded. Then, $\xi_K ({n_1},\ldots,{n_{J}})$ is the event of successful decoding for users $j=1,\ldots,J$ at slots $n_1,\ldots,n_{J}$ and outage for user $K$. Also, $\omega_j ({n_j})$ is the event of successful decoding for the $j$th user in the $n_j$th round and $\phi_K ({n_1},\ldots,{n_{J}})$ is the $K$th user outage event while utilizing the $j=1,\ldots,J$ users'
frequency bands in rounds $n_j+1,\dots,M,j=1,\ldots,J.$ Then, $(a)$ is based on the fact that 1) $\xi_K ({n_1},\ldots,{n_{J}})$'s are disjoint events for different  $n_j$'s, $j=1,\ldots,J,$ 2) different terms of the union are of the same order of $P$ and, thus, 3) at $P\to \infty$ the diversity gain is obtained by considering only one term of the union. Finally, $(b)$ follows from (\ref{eq:rateinr}) at $P\to \infty.$
\end{proof}
Intuitively, the theorem indicates that, at high SNR, e.g., the first $J$ users decode their corresponding messages at their first round, with very high probability. Thus, e.g., the $K$th user can utilize its own $M$ retransmissions and the remaining $J(M-1)$ retransmissions of users $j=1,\ldots,J.$ Consequently, we have $d_K=(J+1)(M-1)+1.$ Then, the diversity gain of the whole system containing $K$ users is given by $\mathop {\min }\limits_{ k=1,\ldots,K }\{d_k\}.$ As an example, with $K=2$ users the diversity gain of the coordinated scheme is increased to $2M-1$, compared to the non-coordinated setting for which we have $d_\text{non-coordinated}=M,$ independently of the number of users. Note that the theorem is presented for the single-antenna setting, while it can be extended for the MIMO setup. Moreover, although the theorem is proved for the INR scheme, the same point holds for the RTD as well (also, see Fig. 2 for examples). Finally, the performance gain is at the cost of coordination overhead mainly at the receivers receiving messages in different frequency slots.
\begin{figure}
\vspace{-5mm}
\centering
  \includegraphics[width=0.97\columnwidth]{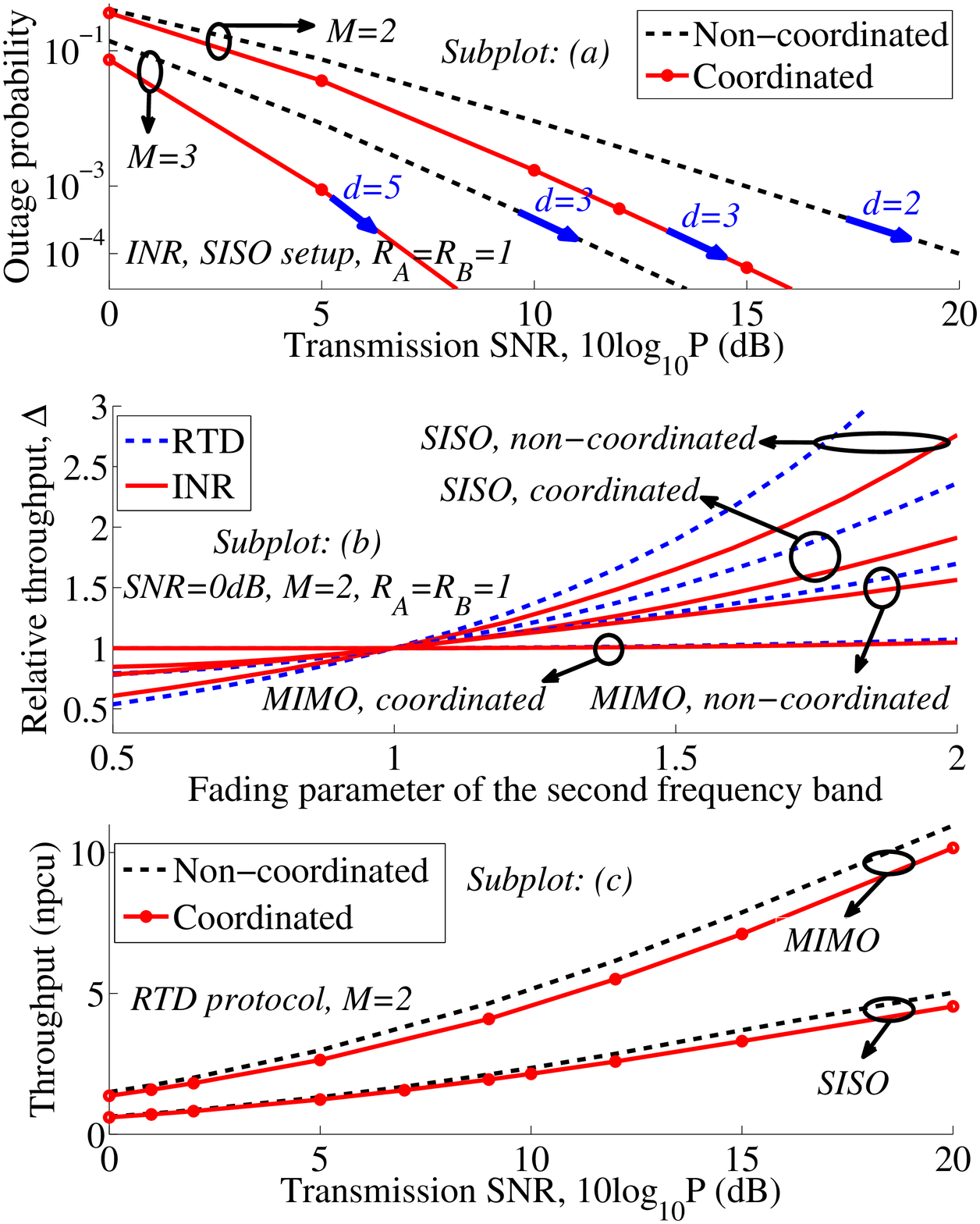}\\\vspace{-5mm}
\caption{Comparison between the coordinated and non-coordinated schemes from (a): outage probability, (b): fairness and (c): throughput perspectives. $K=2,$ $\prescript{}{1}\lambda=\prescript{}{2}\lambda=1$ (except in figure (b), which is for different values of $\prescript{}{2}\lambda$). In figures (a)-(b), $R_\text{A}=R_\text{B}=1.$ In figure (c), the rates $R_\text{A},R_\text{B}$ are optimized in each SNR, to maximize throughput. For the MIMO setup, we set $u=v=2.$}\label{figure111}
\vspace{-4mm}
\end{figure}
\begin{figure}
\vspace{-0mm}
\centering
  \includegraphics[width=0.97\columnwidth]{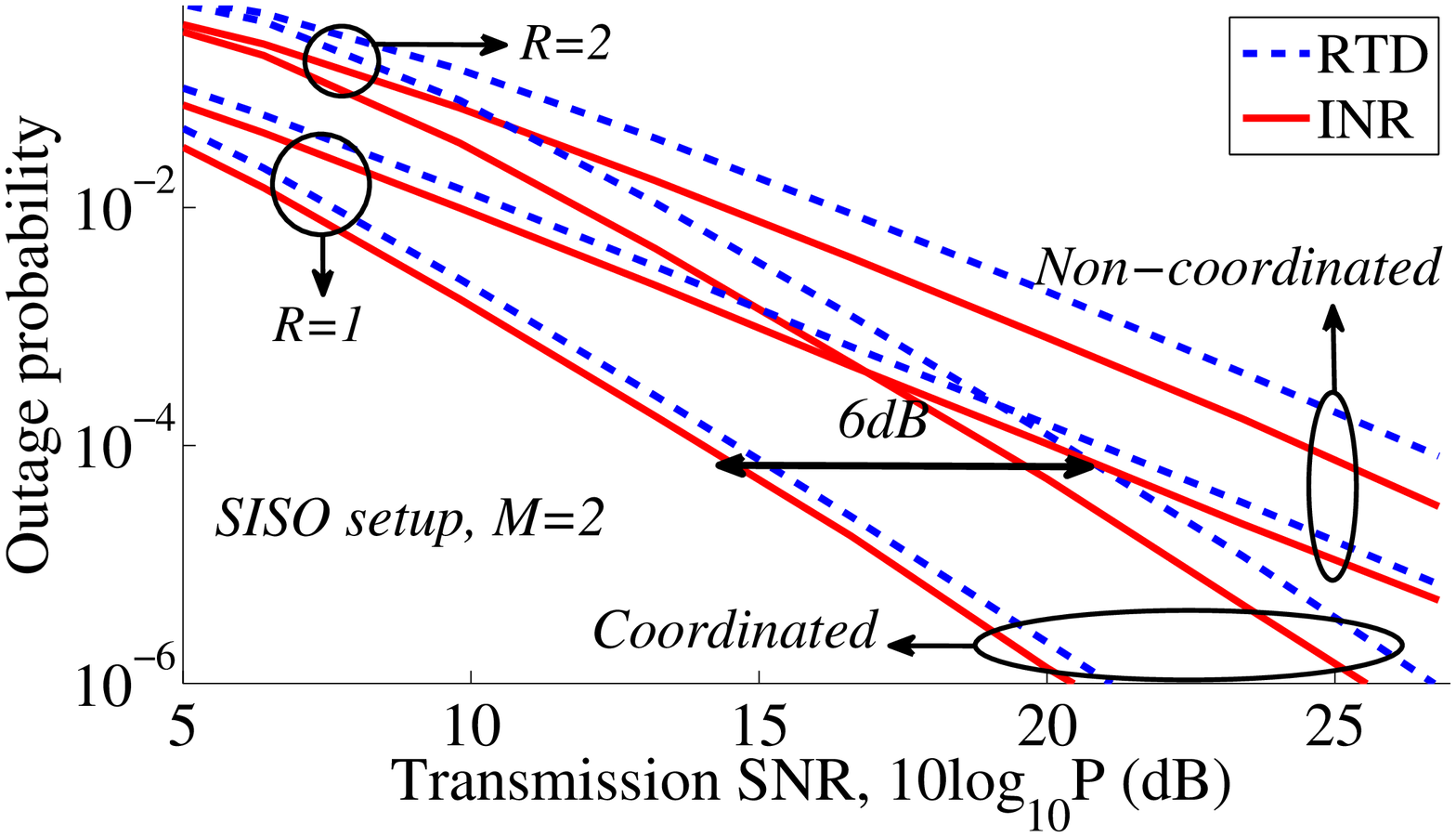}\\\vspace{-4mm}
\caption{Outage probability in the cases with $K=3$ users, SISO setup, $M=2.$ The initial rate of all users is set to $R=1,2.$}\label{figure111}
\vspace{-5mm}
\end{figure}
\vspace{-3mm}
\section{Results and Conclusions}
The simulation results of Fig.1 are obtained for $K=2$ users.
%Here, the probabilities are obtained both numerically and via Theorem 1 which lead to the same outage probabilities.
Here, except for the MIMO setup where the probabilities, e.g., (\ref{eq:rtdMIMO3}) are calculated numerically, the results are obtained both analytically and via Monte Carlo simulations which lead to the same results. Therefore, to avoid too much information in each figure, we plot only one of them. Using the INR, Fig.1a compares the users' outage probability in different schemes. As shown, the coordination decreases the users' outage probability substantially. Also, the impact of coordination on the outage probability increases with the SNR/maximum number of retransmissions $M$. Finally, as shown in the figure, the negative of the slope of the outage probability curves at moderate/high SNRs is the same as the diversity gain derived in Theorem 2. For instance, with $K=2$ and $M=2,$ the diversity gain of the coordinated and non-coordinated schemes are $d=3$ and $d=2$, respectively.

To study the fairness, we plot $\Delta=\frac{\eta_\text{A}}{\eta_\text{B}},$ i.e., the ratio of the users' throughput, for different values of $\prescript{}{2}\lambda.$ Moreover, optimizing the transmission rates by exhaustive search, Fig.1c shows the system throughput (\ref{eq:etaakhar}) for various schemes. As it is seen, the proposed coordinated HARQ scheme improves the users' fairness considerably (Fig.1b), and the throughput loss is negligible in the considered range of SNR (Fig.1c). Also, the users' fairness, outage probability and throughput are improved by increasing the number of transmit/receive antennas.

Setting $\prescript{}{i}\lambda=1,\forall i,$ Fig.2 studies the outage probability in the cases with $M=2$ and $K=3$ users. Here, the initial rate of all users is set to $R=1,2$. Also, if only one user cannot decode its message correctly at the end of round 1, it receives all frequency bands of the second round. Then, in the cases with two unsuccessful users at the end of the first round, the three frequency slots of the second round are randomly allocated to those users such that one of them receives two frequency slots (and the other receives one). As seen, the INR and the RTD schemes have the same diversity gain (see Theorem 2 and its following discussions). Also, the coordination leads to considerable improvements in the energy efficiency. As an example, with $R=1$ and outage probability $10^{-4}$ the coordination improves the energy efficiency of the INR approach by $6$dB.
%the difference between the performance of the RTD and INR schemes increases with the initial rate. However,

To conclude, as demonstrated both theoretically and via simulations, the proposed coordinated HARQ approach leads to considerable users' outage probability and fairness improvement, with limited throughput degradation.
%\section{Conclusion}
\vspace{-1mm}
\vspace{-3mm}

\begin{figure*}[h!]
\vspace{-5mm}
\centering
  \includegraphics[width=0.9\textwidth]{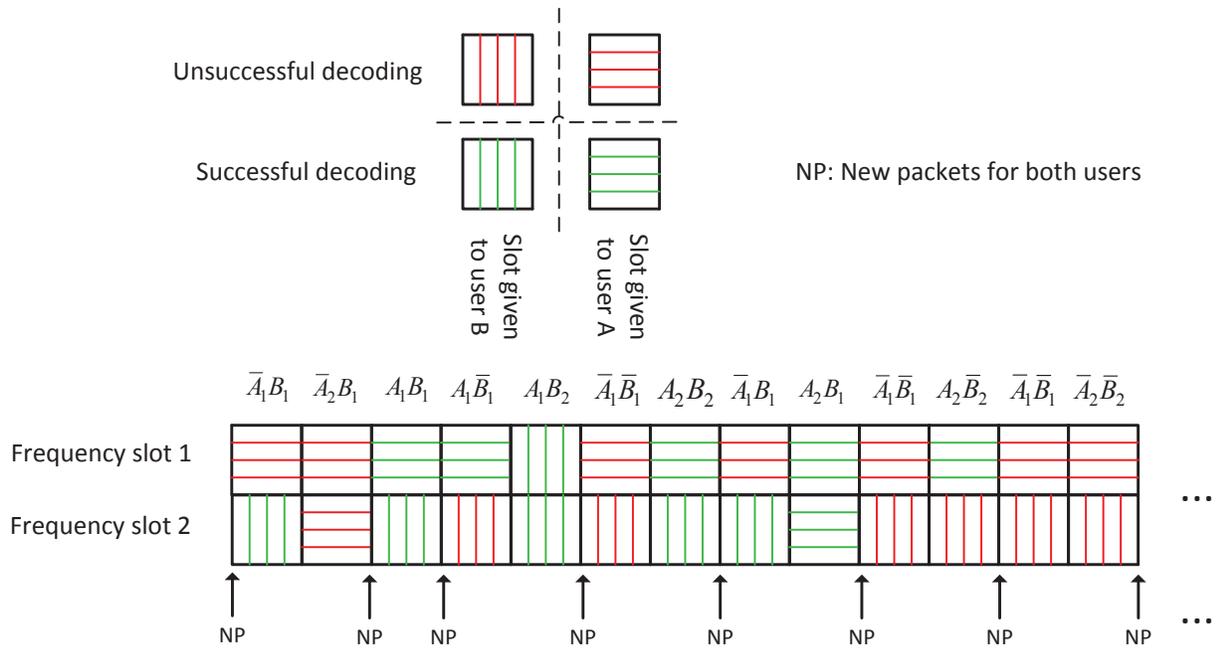}\\\vspace{-5mm}
\caption{Demonstration of the data transmission protocol with $K=2$ users and a maximum of $M=2$ retransmissions. The green (resp. the red) lines indicate successful (resp. unsuccessful) message decoding. Also, the box with horizontal (resp. vertical) lines presents the frequency slot given to user A (resp. user B). Finally, NP stands for new packets for both users. }\label{figure111}
\vspace{-5mm}
\end{figure*}
\bibliographystyle{IEEEtran} %lic.bst is the style file
\bibliography{masterRelay}

%
%
%\begin{figure}
%\vspace{-3mm}
%\centering
%  % Requires \usepackage{graphicx}
%  \includegraphics[width=1\columnwidth]{figRTDspectrumsharingcontinuousTpICC1.eps}\\
%\caption{Secondary channel throughput vs the PU (a) interference power and (b) SINR constraint, continuous communication model, $P_\text{p}=1$.}\label{figure111}
%\vspace{-3mm}
%\end{figure}
%
%\begin{figure}
%\vspace{-3mm}
%\centering
%  % Requires \usepackage{graphicx}
%  \includegraphics[width=1\columnwidth]{figRTDspectrumsharingfigICC3.eps}\\
%\caption{(a): Secondary channel throughput vs outage probability, interference-limited scenario, continuous communication model, $P_\text{p}=1$. Secondary channel throughput vs PU transmission power, SINR-limited scenario.}\label{figure111}
%\vspace{-3mm}
%\end{figure}
%
%\begin{figure}
%\vspace{-3mm}
%\centering
%  % Requires \usepackage{graphicx}
%  \includegraphics[width=1\columnwidth]{figRTDspectrumsharingburstingICC.eps}\\
%\caption{System throughput vs outage probability, (a): interference-limited, (b): SINR-limited scenario, $P_\text{p}=1$. }\label{figure111}
%\vspace{-3mm}
%\end{figure}
%\vspace{-2mm}

\vfill

\end{document}